\documentclass[namedreferences]{SolarPhysics}
\usepackage[optionalrh]{spr-sola-addons} 
\usepackage{graphicx}
\usepackage{url}             

\newcommand{\adv}{    {\it Adv. Space Res.}}
\newcommand{\annG}{   {\it Ann. Geophysicae}}

\newcommand{\apj}{    {\it Astrophys. J.}}

\newcommand{\grl}{    {\it Geophys. Res. Lett.}}

\newcommand{\jastp}{  {\it J. Atmos. Solar-Terr. Phys.}}
\newcommand{\jgr}{    {\it J. Geophys. Res.}}
\newcommand{\mnras}{  {\it Mon. Not. Roy. Astron. Soc.}}

\newcommand{\solphys}{{\it Solar Phys.}}

\newcommand{\ssr}{    {\it Space Sci. Rev.}}

\begin{document}
\begin{article}
\begin{opening}
\title{A Challenging Solar Eruptive Event of 18 November 2003 and the
Causes of the 20 November Geomagnetic Superstorm. IV.~Unusual
Magnetic Cloud and Overall Scenario}
\author{
V.V.~\surname{Grechnev}$^{1}$\sep A.M.~\surname{Uralov}$^{1}$~\sep
I.M.~\surname{Chertok}$^{2}$\sep A.V.~\surname{Belov}$^{2}$\sep
B.P.~\surname{Filippov}$^{2}$\sep V.A.~\surname{Slemzin}$^{3}$\sep
B.V.~\surname{Jackson}$^{4}$}

\runningauthor{Grechnev et al.}

\runningtitle{Magnetic Cloud Responsible for the Superstorm }

\institute{$^{1}$ Institute of Solar-Terrestrial Physics SB RAS,
Lermontov St.\ 126A, Irkutsk 664033, Russia
email: \url{grechnev@iszf.irk.ru} email: \url{uralov@iszf.irk.ru}\\
$^{2}$ Pushkov Institute of Terrestrial Magnetism, Ionosphere and
Radio Wave Propagation (IZMIRAN), Moscow, Troitsk, 142190 Russia
email: \url{ichertok@izmiran.ru} email: \url{abelov@izmiran.ru} email: \url{bfilip@izmiran.ru}\\
$^{3}$ P.N. Lebedev Physical Institute, Leninsky Pr., 53, Moscow,
119991, Russia
email: \url{slem@lebedev.ru}\\
$^{4}$ Center for Astrophysics and Space Sciences, University of
California, San Diego, La Jolla, California, USA email:
\url{bvjackson@ucsd.edu}}

\date{Received ; accepted }

\begin{abstract}
The geomagnetic superstorm of 20 November 2003 with Dst = $-422$ nT,
one of the most intense in history, is not well understood. The
superstorm was caused by a moderate solar eruptive event on 18
November, comprehensively studied in our preceding Papers
I\,--\,III. The analysis has shown a number of unusual and extremely
complex features, which presumably led to the formation of an
isolated right-handed magnetic-field configuration. Here we analyze
the interplanetary disturbance responsible for the 20 November
superstorm, compare some of its properties with the extreme
28\,--\,29 October event, and reveal a compact size of the magnetic
cloud (MC) and its disconnection from the Sun. Most likely, the MC
had a spheromak configuration and expanded in a narrow angle of
$\leq 14^{\circ}$. A very strong magnetic field in the MC up to 56
nT was due to the unusually weak expansion of the disconnected
spheromak in an enhanced-density environment constituted by the
tails of the preceding ICMEs. Additional circumstances favoring the
superstorm were (i)~the exact impact of the spheromak on the Earth's
magnetosphere and (ii)~the almost exact southward orientation of the
magnetic field, corresponding to the original orientation in its
probable source region near the solar disk center.

\end{abstract}
\end{opening}

\section{Introduction}
\label{S-introduction} A series of big eruptive flares in a complex
of large super-active regions 10484, 10486, and 10488 occurred late
in October 2003 (see, \textit{e.g.}, \opencite{Veselovsky2004};
\opencite{ChertokGrechnev2005}; \citeauthor{Gopal2005a},
\citeyear{Gopal2005a,Gopal2005b}; \opencite{Grechnev2005}). These
`Halloween 2003 events' produced geomagnetic superstorms with Dst $=
-353$ and $-383$~nT\footnote{According to the final data of the
Kyoto Dst index service,
\url{http://wdc.kugi.kyoto-u.ac.jp/dst_final/index.html}} on
29\,--\,31 October. A notable event on 18 November occurred in the
decaying active region (AR) 10501 during the second passage across
the solar disk of the former AR~10484. This event produced a yet
stronger superstorm on 20 November 2003 with Dst $= -422$~nT, the
largest one during solar cycle 23 (\textit{e.g.},
\opencite{Yermolaev2005}; \opencite{Gopal2005c};
\opencite{Yurchyshyn2005}; \opencite{Ivanov2006}), and one of the
severest storms in history (probably among top ten in terms of the
Dst index -- see \opencite{CliverSvalgaard2004}).

A number of studies addressed the 18 November 2003 event and its
interplanetary consequences, endeavoring to understand its extreme
geoeffective impact, but its causes remain unclear. The major
outcome of the studies is the oddness of the event, which strongly
deviated from the established correlations between solar and
near-Earth parameters
\cite{Yermolaev2005,Yurchyshyn2005,Srivastava2005,Chertok2013}. In
particular, the magnetic cloud (MC) near the Earth carried an
exceptionally strong magnetic field of about 56~nT, while its
velocity was modest.

\begin{figure} 
\centerline{\includegraphics[width=0.8\textwidth]{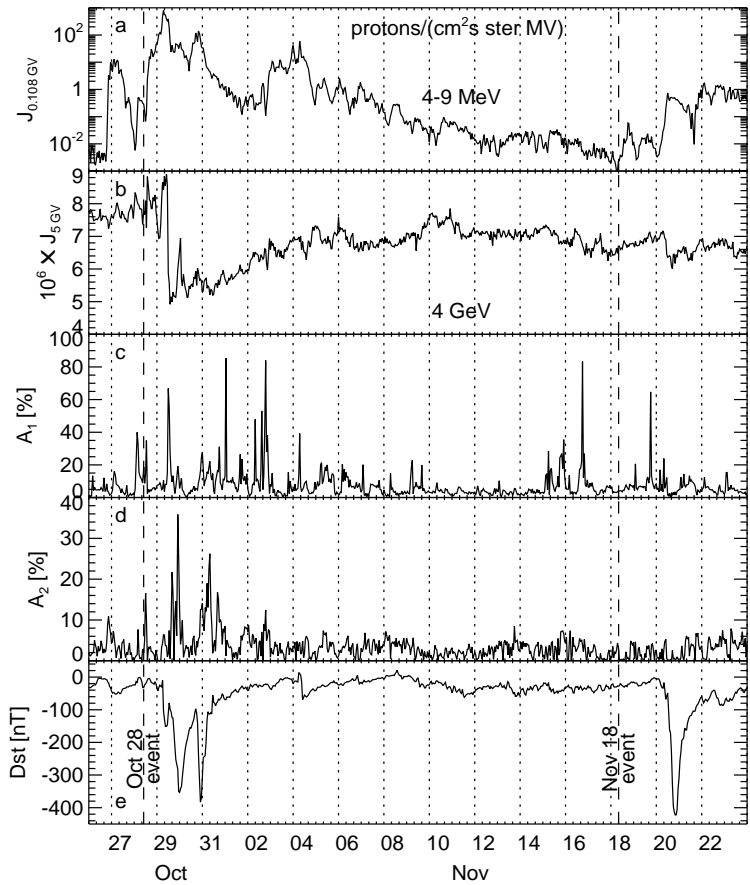}}
\caption{Solar protons (a), galactic cosmic ray data (b\,--\,d,
Courtesy V. Sdobnov), and geomagnetic disturbances (e) during the
late October through November interval. The spherical harmonics of
the galactic cosmic ray pitch-angle anisotropy are shown in panel
c (first harmonic, $A_1$) and d (second harmonic, $A_2$). The
dashed lines mark the 28 October and 18 November solar events. }
\label{F-cosmic}
\end{figure}

The extreme geomagnetic disturbances of 29\,--\,30 October and
20\,--\,21 November (Figure~\ref{F-cosmic}e) were produced by solar
eruptions from nearly the same complex of active regions (which
evolved in the meantime), and therefore one might expect the
18\,--\,20 November event to inherit the properties of the Halloween
events (29\,--\,30 October). However, it looks instead like their
antipode. For example, the 28 October solar event produced large and
fast CME and very strong flare emissions in soft X-rays (exceeded
the GOES saturation level of X17.2), hard X-rays, and gamma rays;
huge radio bursts in microwaves (RSTN radiometers saturated) and up
to submillimeters; strong SEP event (Figure~\ref{F-cosmic}a), and a
ground-level enhancement of the cosmic ray intensity, GLE~65. The 29
October event was similar to the 28 October event. On the other
hand, none of the listed extreme properties were found in the 18
November event. It was medium in soft X-rays ($\lsim\,$M5), hard
X-rays, and microwaves. The enhancement of the near-Earth proton
flux was insignificant (but it could have been reduced due to the
easterly position of the active region on 18 November). The
associated CMEs had a moderate speed, size, and brightness, and did
not exhibit any extreme features. The time intervals between the
solar eruptions and the geomagnetic storm onset/peak times were
19/38~h on 29 October and much longer on 20 November, 48/61~h.

The peculiarities are also related to the MC, which reached the
Earth's magnetosphere on 20 November. The $z$-component of the
magnetic field ($B_z$) in the MC was pointed southward (negative),
although the MCs produced by the eruptions from AR~10484 in October
had positive $B_z$. The magnetic helicity of the MCs responsible for
the Halloween superstorms (from AR~10486) was negative, but it was
positive in the MC of 20 November. Moreover, the orientation and
handedness of the magnetic field in the 20 November MC mismatched
those in its presumable source region AR~10501
\cite{Moestl2008,Chandra2010}.

Thus, the extremely geoeffective event of 18\,--\,20 November has
offered four major problems: (i)~incomprehensible causes of its
extremeness, (ii)~its enigmatic solar source, and perplexing
(iii)~orientation and (iv)~handedness of the magnetic field in the
MC. Despite several ideas proposed in the listed studies to address
these issues, satisfactory explanations are still missing. This
enigmatic event offers the challenge in predictions of geomagnetic
disturbances that can occur unexpectedly and can be destructive for
modern and future high-technology societies.

These circumstances were the major reasons that inspired us to
undertake a comprehensive analysis of this event presented in our
Paper~I \cite{Grechnev2014_I}, Paper~II \cite{Grechnev2014_II}, and
Paper~III \cite{Uralov2014}. The analysis has revealed an extremely
complex, unusual solar eruptive event. None of the observed CMEs was
an appropriate candidate for the source of the MC, being only able
to produce a glancing blow on the Earth due to their rather large
angles with the Sun\,--\,Earth line. However, three different
reconstructions of the MC \cite{Yurchyshyn2005, Moestl2008, Lui2011}
showed its central encounter with Earth.

On the other hand, indications were found of an additional eruption,
which probably occurred rather far from AR~10501 but close to the
solar disk center. The presumable erupted magnetic structure was a
right-handed pair of linked tori moving away from the Sun, and
slowly expanding with a radial (lateral) speed of $\approx
100$~km~s$^{-1}$, \textit{i.e.}, within a narrow angle of $\lsim
14^{\circ}$ (Papers II and III). It is not completely clear into
what structure the couple of tori evolved. Most likely, this
structure was disconnected from the Sun and developed into a toroid
(donut-shaped) or spheromak (spherical, no central hole). Due to its
weak expansion, the earthward direction of propagation, and a
presumably small mass, the probability to detect this eruption in
coronagraph images was very low. Therefore this eluding structure
could have evolved into the MC that reached Earth on 20 November.

We will not list all the ideas proposed previously and follow
instead only those results and suggestions that appear to be
consistent with these findings and promising to shed light on the
issues in question. The major reasons for the superstorm of 20
November 2003 were the strongest magnetic field in the MC (close to
a record value, while its speed was ordinary) and its orientation
($\theta = -(49-87)^{\circ}$; see \opencite{Moestl2008}).
\inlinecite{Qiu2007} demonstrated a quantitative correspondence
between the magnetic fluxes in nine MCs and their solar source
regions. Furthermore, \inlinecite{Chertok2013} analyzed the major
geomagnetic storms (Dst~$<-100$~nT) in solar cycle 23 whose solar
source regions were located with sufficient reliability in the
central part of the disk. They found that the intensity of the
geomagnetic storms, as well as the ICMEs' Sun--Earth transit times,
were mainly governed by the parameters of their solar sources, such
as the total unsigned magnetic flux at the photospheric level within
the post-eruption EUV arcades and dimming regions. For example, the
magnetic fluxes in the solar sources of the strongest geomagnetic
storms in cycle 23 ($\mathrm{Dst} < -200$~nT, the near-Earth
magnetic field $|{\bf B}| > 50$~nT, and large southern $B_z$
component) such as the 14--15 July 2000, 22--24 November 2001,
28--30 October 2003, and 13--15 May 2005 superstorms (see,
\textit{e.g.}, \opencite{Wu2005}; \opencite{Cerrato2012};
\opencite{Manchester2014}) were very large, $(240-870) \times
10^{20}$~Mx (maxwell). The severity of the geomagnetic superstorm of
20 November 2003 appears to correspond to the near-Earth magnetic
field of $|{\bf B}|_{\max} \approx 56$~nT with a large southward
component up to $B_z \approx -45$~nT, while the total unsigned
magnetic flux in the eruption region was as low as $130 \times
10^{20}$~Mx, even including the flare arcade in AR 10501 and all
dimming regions. The only way to get an extremely strong magnetic
field with a modest magnetic flux is a small size of the MC. This
conjecture is consistent with the above-mentioned reconstructions of
the MC, all of which showed both its dimensions in the ecliptic
plane to be $< 0.3$~AU ($< 17^{\circ}$). One more indication is that
the Forbush decrease (FD) on 20 November was much less than that
after the 28 October event. Note also the idea about the compression
of the MC due to the interaction among CMEs
\cite{Gopal2005c,Yermolaev2005}.

The mentioned reconstructions were based on the fitting of specific
configurations such as a torus or cylinder \cite{Yurchyshyn2005,
Moestl2008, Lui2011, Marubashi2012} to the observed rotating
magnetic components in the MC. The fitted components more or less
resembled the observed ones within some limited portions of the
ICME, overlapping with each other, but covering somewhat different
parts of the ICME in different reconstructions. On the other hand,
considerations of the $\phi_B$ and $\theta_B$ angles led
\inlinecite{KumarManoUddin2011} to the idea that the MC occupied a
longer part of the ICME along the Sun--Earth line (see also
\inlinecite{Gopal2005c} and \inlinecite{Marubashi2012}). This
possibility suggests that its magnetic structure might be different
from those considered previously.

To verify these suggestions and shed further light on the enigmatic
18\,--\,20 November 2003 event, we address the corresponding ICME
here. In Section~\ref{S-icme} we revisit \textit{in situ}
measurements of the interplanetary disturbance, analyze ground-based
data on cosmic rays, and consider heliospheric three-dimensional
(3D) reconstructions made from the observations of the \textit{Solar
Mass Ejection Imager} (SMEI; \opencite{Eyles2003};
\opencite{Jackson2004}). Then we discuss the results and their
implications in Section~\ref{S-discussion}. In particular, we
endeavor to understand the probable causes of the superstorm, to
follow the whole presumable chain of events from the solar eruption
on 18 November up to the encounter of the MC with the Earth on 20
November, and to outline possible ways to diagnose such anomalously
geoeffective events.

\section{Properties of the ICME}
\label{S-icme} As mentioned above, the 20 November geomagnetic
superstorm was a conspicuous exception to almost all established
statistical correlations (see, \textit{e.g.},
\opencite{Yurchyshyn2005}; \opencite{Srivastava2005};
\opencite{Chertok2013}). What is surprising is that it also promises
an opportunity to find hints on the causes of the superstorm from
its peculiarities. By comparing the 28 October event and related
interplanetary disturbances with those of 18--20 November, we hope
to understand the incomprehensibly large geomagnetic effect of the
18 November eruption.

\subsection{Interplanetary Data}
\begin{figure} 
\centerline{\includegraphics[width=0.85\textwidth]{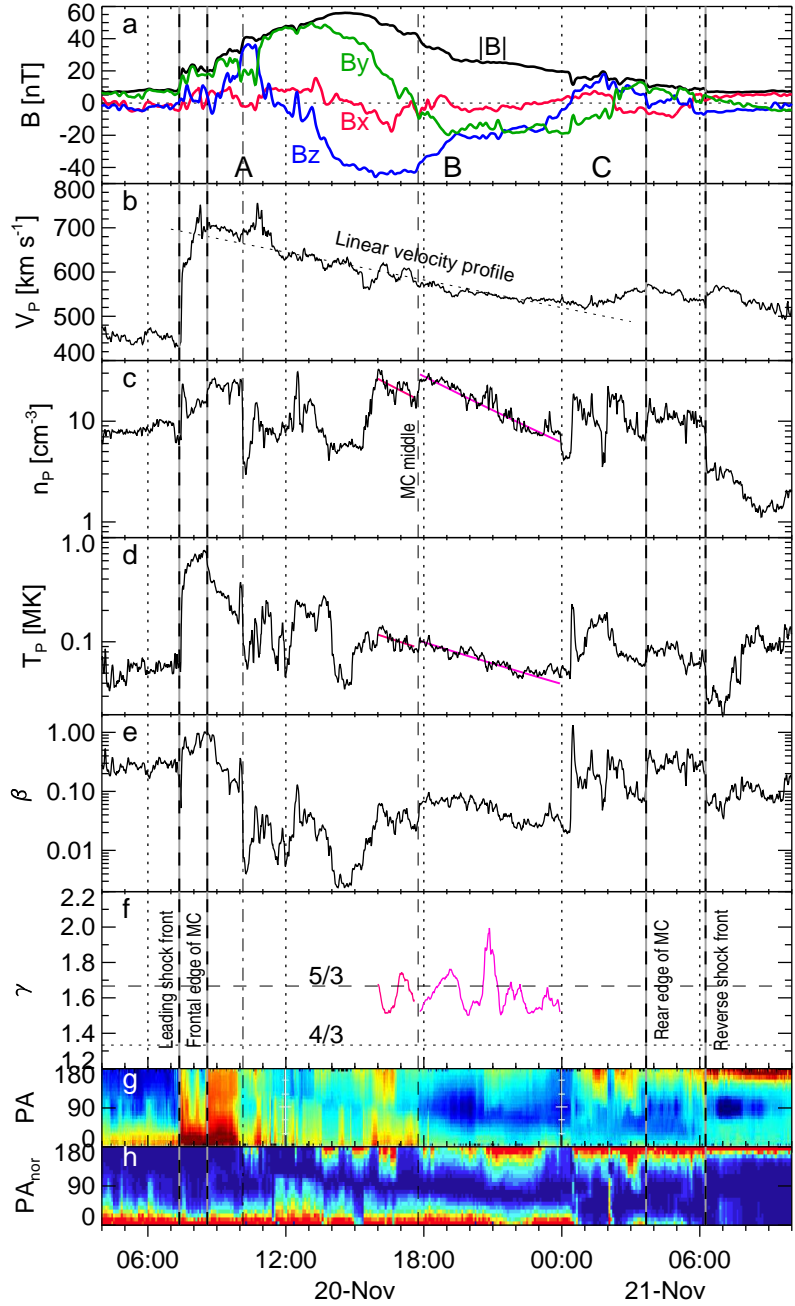}}
\caption{ACE data. (a)~$B_x$, $B_y$, and $B_z$ components of the
magnetic field and its magnitude $|{\bf B}|$. Parameters of the
ejecta: velocity (b), proton density (c) and temperature (d), and
plasma $\beta$ value (e). (f)~Polytropic index $\gamma$ inferred
from the density and temperature in two intervals marked with the
pink and purple lines. (g)~Pitch angle distribution of 272~eV
electrons (minimum blue, maximum red). (h)~Same as (g) but with a
moving normalization (adopted from Marubashi \textit{et al.},
2012).} \label{F-ace}
\end{figure}

Data on \textit{in situ} measurements of the interplanetary
disturbance have been addressed in several studies (\textit{e.g.},
\opencite{Yurchyshyn2005}; \opencite{Yermolaev2005};
\opencite{Gopal2005c}; \opencite{Moestl2008};
\opencite{KumarManoUddin2011}; \opencite{Marubashi2012}).
Nevertheless, some significant particularities in this event were
not discussed previously. Here we consider the near-Earth ICME
observations made with Solar Wind Electron Proton Alpha Monitor
(SWEPAM; \opencite{McComas1998}) and Magnetometer Instrument (MAG;
\opencite{Smith1998}) on \textit{Advanced Composition Explorer}
(ACE). Figure~\ref{F-ace} shows records of the magnetic field
components $B_x$, $B_y$, and $B_z$ along with a magnitude $|{\bf
B}|$ (a), solar wind velocity (b), proton temperature (c) and
density (e), and plasma $\beta$ value inferred from the above
parameters.

The identification of the boundaries of the magnetic cloud was made
as follows. After the arrival of the shock front at about 07:20 UT
on 20 November, the plasma velocity kept on increasing up to the
contact surface that is typical of a bow shock. This fact suggests
significance of the aerodynamic drag. The thick dashed line at 08:35
and the thin dash-dotted line at 10:08 in Figure~\ref{F-ace} mark a
possible arrival time of the MC frontal edge, but this is rather
uncertain. The earlier thick line roughly corresponds to the
position of the highest plasma velocity. Such a situation occurs at
the forward stagnation point of a supersonic body with fixed shape
and size. The later thin line approximately corresponds to the
position where the plasma beta and density underwent a seemingly
sharp drop. This situation is expected for a typical MC. It is
difficult to choose between the two options of the MC leading edge
for the following reasons. On the one hand, the MC must have
expanded with a velocity of the order of 100~km~s$^{-1}$ which was
appreciably higher than the fast-mode speed in the undisturbed solar
wind and would have prevented a stationary flow around the ICME. On
the other hand, the nearly self-similar expansion of a MC that is
often assumed implies conservation of the profile of the plasma beta
formed at the MC creation. As Paper~III showed, the MC hitting the
Earth was probably formed from multiple magnetic structures with
considerably different temperatures, densities, and possibly plasma
beta values. Therefore, sharp changes in plasma beta value without
significant variations in $|{\bf B}|$ in Figure~\ref{F-ace} might
correspond to these different parent structures.

We consider the first thick dashed line as the leading edge of the
MC because of (i)~equality of the $B_z$ flux in a closed magnetic
structure (see below), (ii)~a sharp density increase in suprathermal
electrons (Figure~\ref{F-ace}g) discussed later, and (iii)~sharp
changes in the proton temperature and density. The conditions (i)
and (iii) were also used in identifying the rear edge of the MC.

One more feature deserving attention is a weak reverse shock in
Figure~\ref{F-ace} at 06:10 on 21 November. Its presence as well as
the presence of the forward shock indicates that the expansion of
the MC was not free for all directions. A reverse shock can be due
to two reasons. One is overexpansion of the MC if its earthward bulk
velocity was less than the velocity of its own expansion. This was
not the case here. The second possibility for the development of the
reverse shock appears if the MC was pressed from behind by the
faster flow of the disturbed solar wind. The latter situation is
favored if the MC was disconnected from the Sun. In this situation
the MC would not have been protected from lateral disturbances such
as shocks and related high-speed flow which could come from
expanding CME1, CME2, and CME3. If the magnetic flux rope was
connected to the Sun, then the extended magnetic structure would
have protected the top of the MC from such lateral disturbances.

The trend of the velocity in the MC (region B in
Figure~\ref{F-ace}a) is close to a linear one (dotted line), which
is an attribute of a self-similar expansion \cite{Low1984}. Such
expansion occurs if all forces affecting the ejecta (magnetic
forces, plasma pressure, and gravity; so far we have neglected the
aerodynamic drag because its effect is not self-similar) decrease
with distance by the same factor \cite{Low1982,Uralov2005}. This
condition is satisfied if the polytropic index of the MC gas is
$\gamma = 4/3$. Its actual value in the MC interior can be estimated
from the proton density and temperature (Figures~\ref{F-ace}c and
\ref{F-ace}d) as $\gamma = n/T\, \left( dT/dt \right) / \left( dn/dt
\right) +1$ as shown in Figure~\ref{F-ace}f. This expression is
valid if the entropy distribution inside the MC is uniform.
Averaging of values obtained in this way provides an average
$\gamma$ inside the MC. To suppress fluctuations in the computation
of the derivatives, we fitted the trends within two intervals with
smooth functions (pink and purple). The actual $\gamma$ varies
around $5/3$; the difference from $4/3$ does not seem to be
significant, because $\beta \leq 0.1$ within this interval.
(According to \inlinecite{Farrugia1995}, the estimated value of
$\gamma$ supports the spheromak configuration rather than a flux
rope.) These circumstances allow us to calculate an
expansion-corrected `snapshot' of the ICME assuming its expansion to
be exactly self-similar. We infer the time-dependent expansion
factor $\xi(t)$ from the linear fit to the velocity profile. The
correction factor for the magnetic field strength in the snapshot
related to moment $t_0$ is $[\xi(t_0)/\xi(t)]^2$ (from the magnetic
flux conservation), and the correction factor for the density is
$[\xi(t_0)/\xi(t)]^3$.

\begin{figure} 
\centerline{\includegraphics[width=\textwidth]{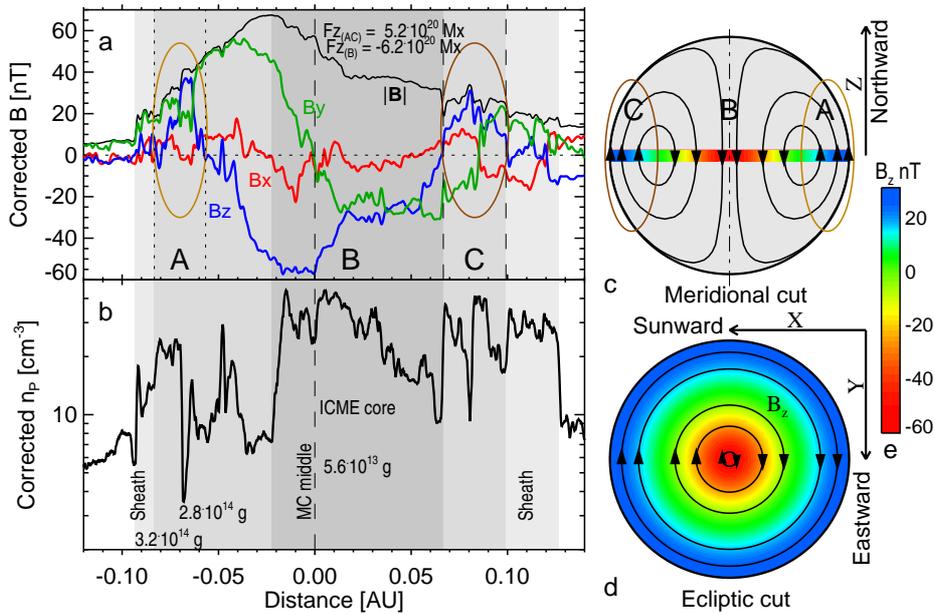}}
\caption{A `snapshot' of the magnetic field (a) and density (b)
distributions in the ICME along the Sun--Earth line according to the
expansion-corrected ACE data. The ICME size corresponds to the first
contact with ACE. (c,\,d)~A magnetic configuration of a perfect
spheromak with its axis shown perpendicular to the ecliptic plane
for simplicity. The brown ovals outline the regions presumably
belonging to the leading (A) and trailing (C) parts of the MC.
(e)~The color scaling of the $B_z$ component in a perfect spheromak
shown in panels (c) and (d) and roughly corresponding to $-B_z$ in
Figure~15 of Lui (2011). } \label{F-ace_cor_spheromak}
\end{figure}

Figures \ref{F-ace_cor_spheromak}a and \ref{F-ace_cor_spheromak}b
show such `snapshots' for the magnetic field and density
distributions in the ICME along the Sun\,--\,Earth line
corresponding to the first contact with the ACE spacecraft. The
expansion factor evaluated from the linear velocity profile in
Figure~\ref{F-ace}b was squared and applied to the magnetic field
strength; likewise it was cubed and applied to the density. The
total length of the ICME in the Sun\,--\,Earth direction was
$\approx 0.22$~AU and, according to the reconstructions of the MC
\cite{Yurchyshyn2005, Moestl2008, Lui2011}, its axis passed close to
the ACE spacecraft.

The behavior of the $B_x$ component, which alternately reached
significant positive and negative values within the MC (see also
Figure~12 in \opencite{Lui2011}), rules out a glancing blow of the
flux rope connected to the Sun; otherwise, the sign of $B_x$ would
not change in the MC. Variations in the magnetic components in
regions A and C outlined with the brown ovals in
Figure~\ref{F-ace_cor_spheromak}a look like their regular
continuations from region B rather than piled-up fluctuations in the
sheath. To verify this conjecture, we estimated the total magnetic
fluxes of the $B_z$ component in region B, on the one hand, and the
corresponding sum in regions A and C, on the other hand, assuming
the ICME cross section to be circular. To take account of the ICME
asymmetry in the Sun--Earth direction, the two estimates for each
ICME half were averaged. The result shown in
Figure~\ref{F-ace_cor_spheromak}a demonstrates that the difference
between the magnetic fluxes in regions B ($-6.2 \times 10^{20}$~Mx)
and A+C ($5.2 \times 10^{20}$~Mx) did not exceed 20\%. These values
are close to the estimate of $5.5 \times 10^{20}$~Mx obtained by
\inlinecite{Moestl2008}, despite the coarseness of our approach.
Hence, the magnetic field variations in regions A and C were, most
likely, not sporadic.

The presence of the opposite magnetic $B_z$ components and the
equality of their fluxes together with the same situation for $B_y$
(while $B_x$ was rather small) indicate their balance in the ICME,
\textit{i.e.}, a closed magnetic field in the MC. This is only
possible if the MC was not connected to the Sun, so that its
configuration was either an isolated toroid or a spheromak. In
either case, the angle between the MC axis and the ecliptic plane
must be large. We adopt the spheromak configuration and later
confirm this assumption by different indications considered in
Section~\ref{S-icme}.

A simplest 2.5-dimensional configuration which satisfies the
equality of magnetic fluxes is an infinitely long cylinder with a
linear force-free magnetic field inside: $B_z = J_0(\alpha r),
B_\varphi = \sqrt{(B_x^2+B_y^2)} = J_1(\alpha r)$ where $J_n$ is the
Bessel function of order $n$. The boundary of the cylinder is
determined by the first root of the equation $J_1(\alpha r) =0$. The
total magnetic flux through the normal cross section of such a
magnetic flux rope is zero. A 3D analog of this situation is a
spheromak whose total magnetic flux equals to zero in both
equatorial and meridional cross sections. The boundary of a
spheromak is determined by the first root of the equation
$j_1(\alpha r) = 0$ where $j_1$ is the spherical Bessel function of
order 1. A transformation of a long cylindric flux rope into a thin
toroid does not significantly change the inherent situation for an
infinitely long flux rope. A force-free flux rope whose small radius
is determined by the condition $J_1(\alpha r) =0$ cannot have its
footpoints in the solar surface. If it were so, then the total
magnetic flux in each photospheric footpoint of the flux rope would
be zero, which obviously disagrees with observations of eruptive
coronal flux rope structures. Thus, the MC was most probably a
force-free self-closed magnetic structure rather than that connected
to the Sun, and the total magnetic flux in its cross section was
zero.

It is important to note that usually the boundary of a MC is
considered as the surface on which the axial magnetic field of
either a loop-like flux rope or toroid is zero. The linear
force-free approximation in this situation corresponds to the
equation $J_0(\alpha r) =0$ whose first root determines the boundary
of a rope. In this case, the total flux of the axial magnetic field
in the flux rope's cross section is maximum but not zero. At present
this scheme is a basis of the technique to fit the internal magnetic
structure of many observed MCs (see, \textit{e.g.},
\opencite{RomashetsVandas2003}; \opencite{MarubashiLepping2007}).
\inlinecite{Marubashi2012} also employed this technique to determine
the geometry of the 20 November 2003 MC. However, this technique
does not permit the presence of the opposite magnetic $B_z$
components with the equality of their fluxes that we emphasize. The
fitting technique usually employed is not usable in the 20 November
2003 event exactly for this reason.

A quite different Grad\,--\,Shafranov reconstruction technique was
used by \inlinecite{Yurchyshyn2005}, \inlinecite{Moestl2008}, and
\inlinecite{Lui2011}. This technique allows one to reconstruct a
two-dimensional cross section of a MC and to evaluate the
inclination of its axis only if the MC is a nearly straight portion
of a magnetic flux rope with the curvature, plasma pressure, and
magnetic field slowly varying along the flux rope axis. Regrettably,
the reconstruction intervals in these papers did not fully cover all
the three intervals A, B, and C shown in
Figure~\ref{F-ace_cor_spheromak}. \inlinecite{Moestl2008} considered
a large portion of interval B starting from its earlier boundary.
\inlinecite{Yurchyshyn2005} included the same portion and a half of
interval A. \inlinecite{Lui2011} considered the whole interval B and
almost the whole interval C. Nevertheless, Figure~15 from
\inlinecite{Lui2011} shows that the direction of the ${\bf B}_x +
{\bf B}_y$ vector is practically constant from the flux rope center
to its periphery despite the change in sign of $B_z$. The ecliptic
cut of the spheromak in Figure~\ref{F-ace_cor_spheromak}d shows the
same situation. A cut of a force-free magnetic cylinder with a
boundary determined by the condition $J_1(\alpha r) =0$ is similar.
However, one should be aware of the fact that the results of the
Grad\,--\,Shafranov reconstruction of the outer part of a MC can be
distorted if all the MC dimensions are comparable which is the case
for a spheromak. Possibly, this circumstance was responsible for the
appearance of an X-point in the reconstruction of
\inlinecite{Yurchyshyn2005}.

Figures \ref{F-ace_cor_spheromak}c and \ref{F-ace_cor_spheromak}d
show two views of a perfect non-distorted spheromak with its axis
perpendicular to the ecliptic for simplicity. The regions outlined
with the ovals correspond to regions A (dark brown ovals) and C
(bright brown ovals) in Figure~\ref{F-ace_cor_spheromak}a. The
passage of the spheromak across the ACE spacecraft is expected to
produce a response that is close to the observed one. The actual
front/tail asymmetry of the magnetic field indicates compression of
the leading half of the MC and stretch of the trailing one,
suggesting a significant role of the aerodynamic drag. Note that the
toroidal (Figure~\ref{F-ace_cor_spheromak}d) and poloidal
(Figure~\ref{F-ace_cor_spheromak}c) magnetic fluxes in the spheromak
are transposed with respect to a torus (connected to the Sun or
disconnected).

For a spheromak configuration it is possible to estimate the total
mass of the ICME. We consider a simple three-layer spheroid
consisting of (i) a central core, (ii) the middle portion, and (iii)
a sheath. The ICME cross section is assumed to be an oval extending
along the Sun-Earth line with an eccentricity of 2 according to
\inlinecite{Moestl2008}. The estimates for the leading and trailing
halves of the MC are averaged. The results are shown in
Figure~\ref{F-ace_cor_spheromak}b for the three layers separately.
The total ICME mass is estimated to be $\lsim 10\times 10^{14}$~g
and, since the sheath mass was acquired by the ejecta on the way to
the Earth, the initial mass of the CME was, most likely, $\lsim
5\times 10^{14}$~g. This estimate is consistent with the conclusion
of Paper~I that the major part, $(2-4) \times 10^{15}$~g, of the
initial mass of the eruptive filament of $(4-6) \times 10^{15}$~g
remained on the Sun.

Figure~\ref{F-ace} shows that the standoff distance between the ICME
piston and the shock ahead was $\approx 0.02$ AU, significantly less
than a typical value of 0.1 AU \cite{RussellMulligan2002}. According
to the formulas given in this reference, the radius of curvature of
the ICME nose in this case must be $\lsim 0.1$ AU for each of the
$X$ and $Y$ ICME dimensions with any Mach number $M > 1$. This
radius of curvature is consistent with the ICME geometry discussed
in the preceding paragraph and the conclusion of
\inlinecite{Vandas1997} related to a spheromak.

The spheromak configuration of the MC addresses the suggestions of
its small size mentioned in Section~\ref{S-introduction}. The length
of the MC along the Sun-Earth line (the $X$-direction) was about
0.2~AU (see Figure~\ref{F-ace_cor_spheromak}). The $Y$-size was
close to the $X$-size, as the reconstructions of the MC presented by
\inlinecite{Yurchyshyn2005}, \inlinecite{Moestl2008}, and
\inlinecite{Lui2011} showed indeed. Considering all the facts, we
have to conclude that the whole spheromak-shaped ICME expanded
within a narrow cone of $\leq 14^{\circ}$. Therefore, the
corresponding CME could appear from behind the occulting disk of
LASCO/C2 at distances $\geq 16R_\odot$, where the Thomson-scattered
light was meager. Such a CME could only be detected with LASCO if
its mass were very large, which was not the case. It is therefore
not surprising that nobody has succeeded in detecting this CME in
LASCO images.

We have concluded that the magnetic configuration of the MC was
closed, \textit{i.e.}, the MC acted as a trap for charged particles.
This suggestion can be verified by considering the pitch-angle
distribution of suprathermal electrons. Manifestations of
bidirectional streams in pitch-angle maps are sometimes considered
as evidence for the connection of a MC to the Sun, as
\inlinecite{Moestl2008} did. However, bidirectional particle streams
are not exclusively produced by mirror reflections at the footpoints
of a magnetic flux rope anchored on the Sun. For example, closed
isolated torus or spheromak configurations are perfect magnetic
traps, and therefore bidirectional particles can also be present in
such configurations. Thus, we agree with the approach of
\inlinecite{Marubashi2012} who revealed the presence of
counter-streaming electrons during more than half a day on
20\,--\,21 November and considered them among indications of a
closed MC structure.

The pitch angle distribution of electrons in the 272 eV channel
calculated by \inlinecite{Marubashi2012} as 5-min averages
normalized to the maximum flux value in each time bin is shown in
Figure~\ref{F-ace}h. To understand the nature of the bidirectional
electron flows, let us consider the records of the $B_x$ and $B_y$
components shown in Figures \ref{F-ace}a and
\ref{F-ace_cor_spheromak}b (see also Figure~12 of
\opencite{Lui2011}). Before the arrival of the MC, the signs of
these magnetic components were $B_x <0, B_y>0$, and after the
departure of the MC $B_x >0, B_y<0$, suggesting that the MC passed
the ACE spacecraft when it crossed a sector boundary of the
heliospheric magnetic field (anti-sunward before the MC and sunward
afterwards; see also \opencite{Ivanov2006}). This orientation
corresponded to small pitch angles of electrons flowing away from
the Sun before the MC and those close to $180^{\circ}$ after the MC
passage as seen in Figure~\ref{F-ace}h. The bidirectional electron
flows might have appeared due to the reconnection of the MC with the
surrounding magnetic fields, which is supported by a nearly linear
trend separating the electron pitch-angle distribution. The trend
implies a gradual change in the predominant reconnection at one MC
edge to the opposite one as the MC passed through the sector
boundary. Thus, the bidirectional electron flows might have been
absent in the MC without reconnection, and therefore should not be
considered in favor of the connection of the MC to the Sun.

An additional indication of the trap-like behavior of the MC can be
revealed from the pitch-angle distribution\footnote{
\url{http://www.srl.caltech.edu/ACE/ASC/DATA/level3/swepam/index.html}}
of the 272 eV electrons with an overall normalization presented in
Figure~\ref{F-ace}g. Region A of the ICME presumably belonged to the
MC. Indeed, the SWEPAM-E plot shows a sharp increase in the electron
density at the leading MC boundary, which we identified, and saw no
drastic changes afterwards.

In contrast to the considered indications of the unusual spheromak
configuration of the 20 November 2003 MC, the Halloween 2003 MC
showed the properties typical of a classical croissant-like flux
rope. The latter event and its solar source were addressed by
\inlinecite{Yurchyshyn2005}.

\subsection{Data on Heavy Relativistic Particles}
The intensity of geomagnetic storms is known to strongly depend on
the parameters (the sign and value of the $B_z$ component, speed) in
a relatively local ICME part interacting directly with the Earth's
magnetosphere (see, \textit{e.g.},
\opencite{TsurutaniGonzalez1997}), while the depth of FD is
determined by global characteristics of an ICME, particularly such
as its magnetic field strength, size, and speed (see, \textit{e.g.},
\opencite{Belov2009}). Figure~\ref{F-cosmic}e shows that the
geomagnetic effect from the 18 November event (in terms of the Dst
index) was stronger than that of the 28 October event, but the FD
from the 20 November interplanetary disturbance was incomparably
moderate with respect to the huge Halloween event
(Figure~\ref{F-cosmic}b). However, such a comparison of FDs in the
two events may not be adequate. In particular, the 29 October FD was
the largest one ever observed with neutron monitors \cite{Belov2009}
and could acquire its value only due to an extremely rare
combination of circumstances.

For comparison with other events we have used a data base on
interplanetary disturbances and FDs created in IZMIRAN
(\citeauthor{Belov2001}, \citeyear{Belov2001},
\citeyear{Belov2009}). The density variations of cosmic rays in
the data base refer to cosmic rays of 10 GV rigidity, which is
close to the effective rigidity for the majority of ground-based
neutron monitors, so that cosmic ray variations around this
rigidity can be evaluated with a highest accuracy. The FD
magnitude determined in this way was 28.0\% for 29 October FD and
4.7\% for 20 November. The latter magnitude was probably
underestimated and needs correction for the following reasons. An
unusually large magnetospheric variation in the cosmic ray
intensity observed with ground-based detectors on 20 November
resulted in a decrease in the geomagnetic cutoff rigidities at
observation stations \cite{Belov2005} which reduced the FD depth
evaluated over the worldwide neutron monitor network. Thus, the
real magnitude of the 20 November FD was most likely 6\,--\,7\%;
this has been really confirmed by the most recent estimate of
6.6\% by taking account of the magnetospheric effect. This effect
is generally large \cite{Belov2001}, but does not seem to
correspond to the parameters of the associated ICME. With a
strength of the total interplanetary magnetic field $|{\bf B}|$ of
up to 55.8 nT (hourly averages), one might expect a considerably
larger FD, because such a strong magnetic field was observed only
twice over the whole history of the solar wind observations. The
two other events that occurred in November 2001 had the
corresponding FDs with considerably larger amplitudes, 9.2\% and
12.4\%. All but one (31 March 2001) event with strongest magnetic
field exceeding 45 nT caused FDs larger than that of 20 November
(pronouncedly larger, as a rule).

Besides the magnetic field strength in an ICME, the solar wind speed
should be taken into account in these comparisons. Unlike the
magnetic field, the solar wind speed on 20 November was ordinary and
increased only up to 703 km~s$^{-1}$ (hourly data). The FD magnitude
is known to correlate well with the product of the observed maximum
magnetic field strength $|{\bf B}|_{\max}$ and the solar wind speed
$V_{\max}$ (\textit{e.g.}, \opencite{Belov2009}). In the 20 November
2003 event, the $|{\bf B}|_{\max}V_{\max}$ product normalized to 5
nT and 400 km~s$^{-1}$ is 19.6, while the whole range of the product
over the data base is from 0 to 29.6. We have considered all the
events with $|{\bf B}|_{\max}V_{\max} > 15$ that were not influenced
by preceding events. Almost all the events (except for the mentioned
31 March 2001 event) produced deeper FDs than the 20 November 2003
one, with an average value of $10.1 \pm 1.8\%$. This value is
probably underestimated, because data on the solar wind were
incomplete or absent during several largest FDs. Due to this reason,
our sample does not include such FDs as those in August 1972 (25\%)
and in October 2003 (28\%).

Thus, the ICME parameters observed on 20 November 2003 suggested
expectations of a larger FD. This fact implies that a smaller size
and/or an unusual structure of the CME in question can be a reason
for its lowered ability to modulate cosmic rays. It is possible that
two kinds of powerful CMEs/ICMEs exist whose different structure and
other properties produce the difference between their influence on
cosmic rays even with equal parameters of the plasma velocity and
magnetic field strength.

Both the geomagnetic effect and FD of an ICME are considered to
depend on parameters of a MC with the FD depth being independent of
$B_z$; thus, the inconsistency between the geomagnetic effect and
the FD depth provides further support to the small size of the 20
November MC. Indeed, its extent along the Sun\,--\,Earth line was
about a factor of three smaller than that of the Halloween MC, and
reconstructions of the MC \cite{Yurchyshyn2005,Moestl2008,Lui2011}
show its perpendicular size in the ecliptic to be even less than the
Sun\,--\,Earth extent. Thus, the cross-sectional area of the 20
November MC was one order of magnitude smaller than that of the 28
October MC.

An additional indication can be revealed from the pitch-angle
anisotropy of relativistic protons. Their gyroradius is larger than
that of suprathermal electrons by a factor of $\approx 10^5$, which
considerably reduces effects on their pitch-angle anisotropy due to
factors irrelevant to trapping. Therefore, anisotropy of
relativistic protons promises still more reliable indication of a
large-scale magnetic trap connected to the Sun than low-energy
electrons do. Anisotropy appears in ground-level enhancements of
cosmic ray intensity (GLE events) and due to modulation of galactic
cosmic rays by magnetic clouds and corotating interacting regions in
the solar wind.

The spherical harmonics of the pitch-angle anisotropy computed by
\inlinecite{Dvornikov2013} from the data of the worldwide neutron
monitor network using the method of
\citeauthor{DvornikovSdobnov1997}
(\citeyear{DvornikovSdobnov1997,DvornikovSdobnov1998}) are shown in
Figures \ref{F-cosmic}c (first harmonic, $A_1$) and \ref{F-cosmic}d
(second harmonic, $A_2$). The percentage in the figure shows the
excess of the cosmic ray intensity over the opposite direction
(100\% correspond to a two-fold excess). The second harmonic of the
pitch-angle distribution of 4.1~GeV protons indicates that their
trapping was absent just before 20 November, while the first
harmonic was well pronounced, unlike the end of October, when both
harmonics were distinct.

\inlinecite{Richardson2000} showed that significant increases in the
second harmonic of the cosmic ray anisotropy corresponding to
bidirectional particle flows are typical of MCs of large ICMEs. An
example is shown by the ICMEs of 29 and 30 October in
Figure~\ref{F-cosmic}. In addition to the incompatibly moderate FD
with respect to very large Dst, the absence of any increase in the
second harmonic on 20 November clearly indicates that this ICME was
unusual.

\subsection{Heliospheric 3D Reconstructions from SMEI Observations}
While the CME of interest is not detectable in SOHO/LASCO images,
now we consider 3D reconstructions of heliospheric disturbances made
from white-light Thomson scattering observations with SMEI. Three
SMEI cameras allow us to achieve a combined $\approx 160^{\circ}$
wide field of view at a sufficiently high spatial resolution. The
tomographic reconstructions have been made by the Center for
Astrophysics and Space Sciences in University of California, San
Diego (CASS/UCSD).

\begin{figure}    
\centerline{\includegraphics[width=0.995\textwidth,clip=]{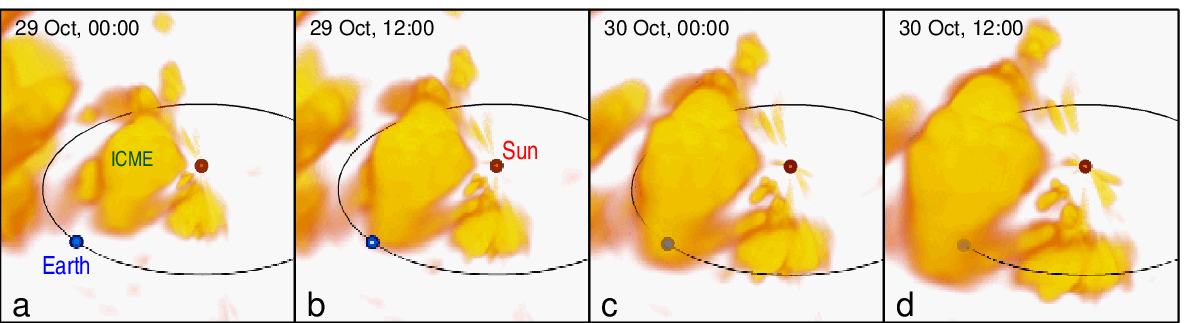}}
\centerline{\includegraphics[width=0.995\textwidth,clip=]{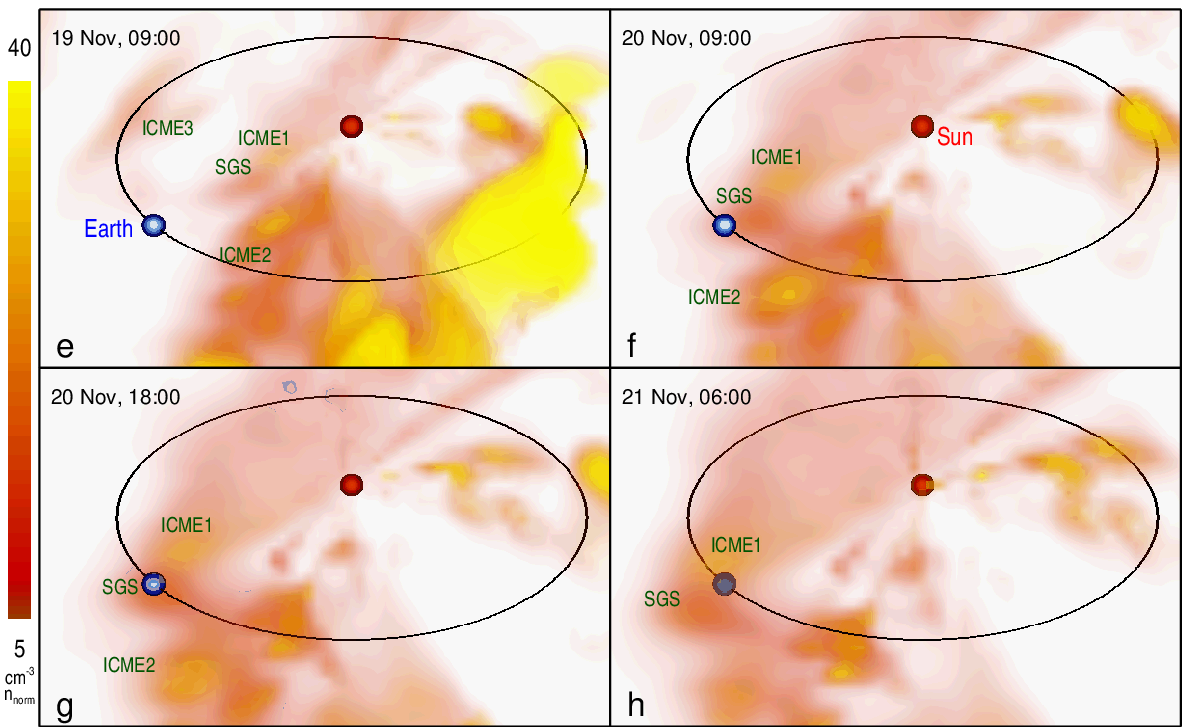}}
\centerline{\includegraphics[width=0.995\textwidth,clip=]{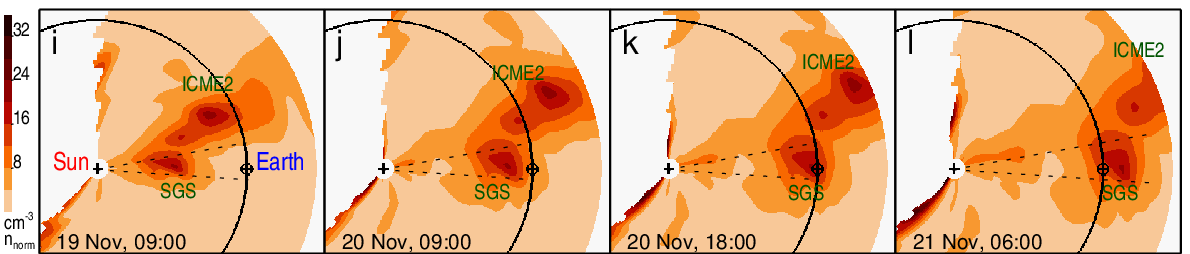}}
\centerline{\includegraphics[width=0.995\textwidth,clip=]{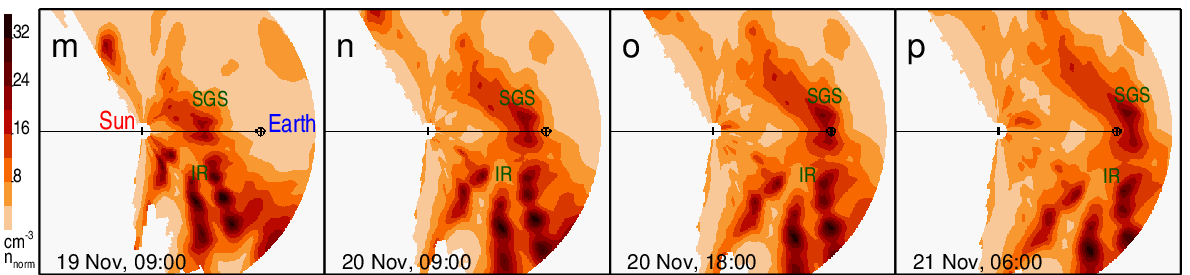}}
\caption{SMEI 3D reconstructions made by the CASS/UCSD team.
(a\,--\,d)~Large croissant\-shaped ICME observed on 29\,--\,30
October suggesting a flux rope connected to the Sun (from Jackson
\textit{et al.}, 2006). (e\,--\,h)~Ejecta observed on 19\,--\,21
November: ICME1, ICME2, ICME3, and a probable compact source of
the geomagnetic storm (SGS). (i\,--\,l)~Ecliptic and
(m\,--\,p)~meridional cuts corresponding to the remote views
(e)\,--\,(h). Here `IR' means a probable intrusion region of CME2
into CME1. The scale bars on the left quantify the densities in
the corresponding rows. } \label{F-smei}
\end{figure}

Figure~\ref{F-smei} presents 3D reconstructions from SMEI data of
the heliospheric response for two ICMEs ejected on 28 October (upper
row) and 18 November 2003, which we want to compare. The images show
heliospheric plasma density distributions as viewed from 3\,AU,
$30^{\circ}$ above the ecliptic plane and $\approx 45^{\circ}$ west
of the Sun--Earth line. The location of the Earth is indicated by a
blue circle with the Earth's orbit viewed in perspective as an
ellipse. The Sun is indicated by a red dot. An $r^{-2}$ density
gradient is removed.

Figures~\ref{F-smei}a\,--\ref{F-smei}\,d adopted from
\inlinecite{Jackson2006} show four successive times of the
heliospheric response to the 28 October CME. The images reveal a
large croissant-shaped ICME expanding from the Sun and suggest
connection of the flux rope to the Sun. The overall picture shows a
presumably typical scenario: a large expanding flux rope reached the
Earth on 29 October and caused (with its southern $B_z$) the severe
geomagnetic storm.

The SMEI 3D reconstructions for the 18\,--\,20 November event in
Figures~\ref{F-smei}e\,--\ref{F-smei}\,h present a very different
picture, which is more complex (see also the movie
\url{SMEI_19-20_11_2003.mpg} accompanying the electronic version of
the paper). To make it clearer,
Figures~\ref{F-smei}i\,--\ref{F-smei}\,l also show the corresponding
ecliptic cuts, and Figures~\ref{F-smei}m\,--\ref{F-smei}\,p show the
meridional cuts. A comparison between the heliospheric disturbances
within a sector of $\approx 120^{\circ}$ embracing the Earth, on the
one hand, and the CMEs observed by LASCO on 18 November by taking
account of their directions and speeds (see Paper~II), on the other
hand, allows one to presumably identify visible ICMEs with the
southeast CME1 and southwest CME2 as was addressed in Paper~II. They
are labeled ICME1 and ICME2. An extended density enhancement labeled
ICME3 in Figure~\ref{F-smei}e appears to correspond to the fastest
far-east CME3.

A far-west Y-like feature resembles in shape the darkening observed
with CORONAS-F/SPIRIT at 304~\AA\ (see Paper~I). However, the speed
of the 304~\AA\ darkening was only $\lsim 100$~km~s$^{-1}$, which
rules out their association.

A large inhomogeneous density enhancement (IR) south of the
ecliptic, which is visible in the meridional cuts
(Figures~\ref{F-smei}m\,--\ref{F-smei}\,p) as a multitude of small
blobs, appears to correspond to the region where CME2 intruded into
CME1 (see Paper~II). The intrusion region appears to have missed the
Earth being south of it.

A feature of our major interest is a rather compact blob exactly
impacting the Earth, in which is a moderate density enhancement
presumably associated with the source of the geomagnetic storm
(SGS). The passage of the blob across the Earth temporally
corresponded to the ACE measurements (Figure~\ref{F-ace}c). The
maximum density in the lower-resolution SMEI two-dimensional cuts
passed through the Earth at about 00 UT on 21 November
(\textit{i.e.}, between Figures \ref{F-smei}k and \ref{F-smei}l)
which reasonably corresponds to the weighted center of the
expansion-corrected higher-resolution ACE record in
Figure~\ref{F-ace_cor_spheromak}b (before the density minimum at the
end of the ICME core, \textit{cf.} Figure \ref{F-ace}c). The SGS
dimensions correspond to expectations: the dotted lines in
Figures~\ref{F-smei}i\,--\ref{F-smei}\,l delimit a sector of
$14^{\circ}$ where the bulk of the ICME mass was concentrated. The
images do not leave any doubt that just this blob was associated
with the MC responsible for the geomagnetic disturbance.

The density in the blob was moderate, peaking at 22~cm$^{-3}$
according to the SMEI tomography, which is close to the ACE
measurements. The blob was surrounded by a larger enhanced-density
cloud suggesting a possible influence from the southwestern ICME2
and southeastern ICME1 as well as the southern intrusion region. As
all the three representations of the SMEI reconstructions show, the
shape of the blob suggest neither a larger isolated torus nearly
perpendicular to the ecliptic plane nor a croissant connected to the
Sun. The spheromak configuration appears to be most appropriate.
Thus, the SMEI data put the last missing point in hunting out the
mysterious source of the 20 November superstorm.

\section{Discussion}
\label{S-discussion} All the observational facts and indications
considered in Section~\ref{S-icme} lead to the conclusion that the
20 November MC was a spheromak of a small size disconnected from the
Sun. These peculiarities of the 20 November ICME must be reflected
in its propagation between the Sun and Earth. Let us compare the
corresponding properties of the 18 November and 28 October ICMEs.

\subsection{Propagation of the 18 November and 28 October ICMEs}
The average velocity of the 18\,--\,20 November ICME between the Sun
and Earth was about 865~km~s$^{-1}$, while its velocity at the first
contact of the ICME body with ACE was $v_1 \approx 700$~km~s$^{-1}$.
This suggests its considerable deceleration. Assuming that the
disconnected 18\,--\,20 November ICME moved almost in the free-fall
regime, it is possible to estimate its initial velocity $v_0$ from
energy conservation, $v_0 = \sqrt{v_1^2 + 2 G M_\odot\left[1/R_\odot
- 1/(1\ \mathrm{AU}) \right]} \approx 930$~km~s$^{-1}$ (the initial
velocity could be still higher, if the aerodynamic drag was
efficient). A conspicuous deceleration of the 20 November ICME
supports its disconnection from the Sun.

The situation was different for the ICME that erupted on 28 October
at about 11~UT and whose body reached ACE on 29 October at about
08~UT according to the ACE/SWICS and ACE/MAG data (ACE/SWEPAM did
not operate because of a large particle event). The average transit
velocity of the ICME was $\approx 1980$~km~s$^{-1}$, while the
frontal speed of the ICME body slightly exceeded 1900~km~s$^{-1}$
(ACE/SWICS, He$^{++}$ bulk speed). Thus, the deceleration of the
28\,--\,29 October ICME was inconspicuous as if it also moved in the
free-fall regime (in this case, $v_0 \approx 2000$~km~s$^{-1}$)
despite its huge size and a much higher speed relative to the
18\,--\,20 November ICME. With such a high speed the gravitational
deceleration is as small as 5\% of the speed so that it is not
possible to reveal it reliably. However, significant deceleration of
the 28\,--\,29 October ICME due to strong aerodynamic drag could be
expected. This apparent inconsistency can be accounted for by a
toroidal propelling force of the 28\,--\,30 October flux rope
connected to the Sun. The toroidal propelling force might have
decreased dramatically as the expansion of the ICME, but its
influence probably was able to ensure the observed velocity of the
ICME, while the actual $v_0$ was probably higher.

Some properties of the 20 November MC were atypical of magnetic
clouds. The inhomogeneous temperature distribution in the MC
(Figure~\ref{F-ace}d) provides a hint that the MC was possibly
formed from structures with different temperatures, as we have
concluded in Paper~III.

The inclination of the 20 November MC axis to the ecliptic estimated
by different authors ranged within $\theta = -(49-87)^{\circ}$ (see
\opencite{Moestl2008}), which is reasonably close to the orientation
of the dipole at the eruption site, $\alpha = -80^{\circ}$, as was
evaluated in Paper~III. Thus, the orientations of the dipole on the
Sun from which the spheromak was formed and the MC near the Earth
were close to each other, and no significant rotation of the ejecta
was required (which was among the problems discussed by
\opencite{Moestl2008}).

The magnetic flux conservation along with the almost exactly
southward orientation of the magnetic axis explains the extreme
geomagnetic effect of the slowly expanding MC in the ICME. The
causes of the unusually slow expansion of the ICME deserve further
study. We assume they might be due to the following reasons.
\begin{itemize}
\item
The major condition for the slow expansion of the ICME was its
disconnection from the Sun. Otherwise, at least, two of ICME
dimensions must be $\geq 1$~AU near the Earth.
\item
The maximum magnetic field strength in a spheromak is $\approx 20\%$
higher than in a cylinder or thin torus under the same outer
pressure at their boundaries.
\item
The inherent ICME expansion could be restrained by the
enhanced-density environment, almost fourfold with respect to
nominal conditions. This dense environment could be due to combined
effects of the tail of the preceding CMEs, the lateral pressures
from ICME2 and ICME1 (including the intrusion region) and, possibly,
from the flank of the shock ahead of ICME3.
\item
The orientation of the magnetic axis of the spheromak nearly
perpendicular to the direction of its motion made the drag pressure
at its surface to preserve the speromak configuration and to prevent
its transformation into a toroid expected for a freely expanding
spheromak \cite{Vandas1997}.
\end{itemize}
The combined effect of the last two factors disfavored a free
expansion of the ICME.

\subsection{Why the 18\,--\,20 November 2003 Event Hindered Understanding?}
The analysis presented in our Papers I\,--\,IV appears to reconcile
all the challenges of the 18\,--\,20 November 2003 event. Moreover,
now it seems to be strange that some important aspects of the event
were not revealed previously. For example, there were several
indications of the small size of the MC: incomparable Dst value and
the depth of the FD; the extremely strong magnetic field despite of
the suggestion of an ordinary magnetic flux (\textit{cf.}
\opencite{Qiu2007}); the small reconstructed cross section of the MC
\cite{Yurchyshyn2005, Moestl2008, Lui2011}. Note that the reduced MC
size also corresponds to the initial idea of
\inlinecite{Yermolaev2005} and \inlinecite{Gopal2005c} about the
compression of the MC by other CMEs.

The spheromak configuration could be best recognized by considering
its entire magnetic field distribution, without ignoring the
positive $B_z$ portions in the leading and trailing portions of the
ICME. Then it is preferred from the magnetic field records that the
orientation of the spheromak was nearly perpendicular to the
ecliptic, although other spheromak configurations of MCs have been
extensively considered for a long time (\textit{e.g.},
\opencite{Ivanov1985}; \opencite{Farrugia1995};
\opencite{Vandas1997}; \opencite{Shiota2010}; \opencite{Zhou2012};
and many others). One of the lessons of the 20 November MC is that
an idealized consideration of a MC as a smooth `magnetic reservoir'
of low-proton-temperature plasma is not always justified.

The comparison of the 3D reconstructions from SMEI white-light
observations in Figures~\ref{F-smei}f and \ref{F-smei}g with those
made from Ooty observations of interplanetary scintillations (IPS)
presented by \inlinecite{KumarManoUddin2011} in their Figure~16
leaves an impression that important ICMEs could also be recognized
in those images, although they do not look as clear as the SMEI
reconstructions. The Ooty IPS images also show suggestions of a
compact blob hitting the Earth, a tail of the southwestern ICME2
and, possibly, the southeastern ICME1 as well as the intrusion
region (IR) in the velocity images. The conclusion of these authors
would have possibly been different from the idea about the two
merged CMEs, if they paid more attention to the indications of a
small size of the MC; 3D reconstructions for a longer time interval
would also be probably useful.

\inlinecite{Moestl2008} suspected the mismatch between the
handedness of the MC and the presumed solar source region AR~10501.
\inlinecite{Chandra2010} have convincingly confirmed this
conjecture. This mismatch, along with the conclusion of
\inlinecite{Grechnev2005} that the eruptive filament had not left
the Sun, seems to be sufficient to warn against a simple scenario in
which the MC is considered as a stretched magnetic rope initially
associated with the pre-eruption filament in AR~10501. However, this
circumstance was not fully acknowledged. Instead,
\inlinecite{Chandra2010} proposed a right-handed eruption from a
small part of AR~10501, although the CME onset times estimated by
\inlinecite{Gopal2005c} did not support this conjecture. The
cautious suggestion of \inlinecite{Chandra2010}, \textit{``Should
this injection} [of the positive helicity flux] \textit{occur over
six days at the same rate an accumulated helicity of the order of
$10^{26}$ Wb$^2$ will be injected, enough to explain the helicity
carried by the positive MC''}, seems to be too extreme an
interpretation. We would like to point out, however, that one of the
severest storms in history was attributed to a partial eruption from
a minor region.

Our caution is also applied to the study by
\inlinecite{Marubashi2012} who presented an undoubtedly valuable
method aiming at a general understanding how the encounter of a MC
with the Earth occurs. Their consideration of the whole sequence of
events starting from the solar eruption on 18 November was based, in
particular, on the assumptions of (i)~the correspondence between the
MC and the eruption region in AR~10501 in handedness and orientation
of the magnetic field, (ii)~the constancy of the direction of the
axial magnetic field in the MC, and (iii)~the association of the MC
with CME2 or, less probable, with CME1. None of these assumptions
was confirmed. Moreover, because the authors ignored the
positive-$B_z$ regions in the MC (like all other researchers), their
consideration of its magnetic configuration could not be perfect.

\subsection{Expansion Factor}
The fact that the geomagnetic superstorm of 20 November 2003 was
produced by the ICME with a total magnetic field up to $|{\bf
B}|_{\max} \approx 56$~nT and a southward $B_z$ up to $-45$~nT does
not contradict the well-known patterns of events (\textit{e.g.},
\opencite{Burton1975}; \opencite{WuLepping2005}). Other solar cycle
23 geomagnetic superstorms mentioned in Section~\ref{S-introduction}
were also caused by the ICMEs with magnetic field strengths of
$|{\bf B}|_{\max} > 50$~nT. The 20 November 2003 ICME had even a
considerably lower speed ($\approx 700$~km~s$^{-1}$) than the other
ICMEs responsible for the superstorms (up to $ 1900$~km~s$^{-1}$).
Therefore, the key to the extremeness of the 18--20 November event
was primarily related to the retention of the strongest magnetic
field up to the Earth orbit.

Thus, one more important outcome of our analysis is a significant
role of the expansion factor for the geoeffectiveness, along with
the strength and orientation of the magnetic field in a MC as well
as its speed. As \inlinecite{Chertok2013} suggested, the magnetic
field and speed of a MC are largely determined by the magnetic flux
in the eruption region; see also \inlinecite{Qiu2007}. The magnetic
flux conservation leads to an estimate of the total magnetic field
in a MC of $|B_{\mathrm{MC}}| \approx B_0 / (L_{\mathrm{MC}}/L_0)^2$
where $L$ is the size and the `0' indices designate the solar source
region. While the sign and value of the $B_z$ component can reduce
the geomagnetic effect in terms of the utmost possible situation of
$B_z \approx -|B|$, an unusually small expansion factor can
considerably strengthen the geomagnetic impact. The expansion factor
can be reduced either due to larger $L_0$ (typical of quiescent
filaments) or due to smaller $L_{\mathrm{MC}}$, which most likely
was the case of 20 November 2003. The range of variations of the
squared expansion factors $(L_{\mathrm{MC}}/L_0)^2$ can probably
exceed a factor of ten.

We considered a large intensity of the geomagnetic storm along with
a moderate FD as an indication of the small size of the MC, which
turned out to be justified in the 20 November 2003 event. A similar
anomaly was also the case in the major geomagnetic storms of 31
March 2001 (Dst $= -387$~nT, FD of 4.1\%) and of 8 November 2004
(Dst $= -374$~nT, FD of 5.2\%). The latter storm was followed by
another one on 10 November 2004 (Dst $= -263$~nT, FD of 8.3\%), for
which the Dst \textit{vs.} FD anomaly was not as challenging. It is
possible to check the conjecture of the small size of the ICME for
the November 2004 events by looking at SMEI reconstructions (SMEI
observed during 2003\,--\,2011) available at
\url{http://smei.ucsd.edu/smeidata.html} and
\url{http://smei.ucsd.edu/test/index.php?type=smei3drecons}. Indeed,
on 8\,--\,10 November 2004 they showed two earthward ICMEs of a
moderate size following each other and did not suggest their
connection to the Sun.

The contribution of the expansion factor to the geoeffective
importance of a solar eruption can probably be responsible for an
additional scatter in already loose correlations between parameters
of solar eruptions and space weather disturbances (see,
\textit{e.g.}, \opencite{CliverSvalgaard2004};
\opencite{Yurchyshyn2005}; \opencite{Chertok2013}). The expansion
factor, which can be especially reduced for disconnected MCs, might
have probably contributed to such abnormal events as the 13\,--\,14
March 1989 superstorm (Dst $= -589$~nT) and the one after the
Carrington event on 1 September 1859 (\opencite{Carrington1859};
\opencite{Tsurutani2003}); the estimated Dst value was of the order
of $-850$~nT according to \inlinecite{SiscoeCrookerClauer2006}.
Previous studies on such events (\textit{e.g.},
\opencite{Tsurutani2003}; \opencite{CliverSvalgaard2004}) did not
consider this possibility, while it seems to be qualitatively clear
what could occur if an event similar to the 18 November 2003 one
involved an eruption in much stronger magnetic fields like the 28
October 2003 event.

\subsection{Overall Scenario of the Event}
Although the whole event was extremely complicated, a key to its
overall scenario seems to be comprehended now. The eruption of the
left-handed main filament from AR~10501 was followed by its
collision with a topological discontinuity in a coronal null point
(whose projection was close to the solar disk center) on its way
out. The eruption resulted in (i)~disintegration of the filament,
which transformed into an inverse-Y-like cloud flying along the
solar surface and probably eventually landed on the Sun, (ii)~a
forced eruption of CME2, and (iii)~reconnection of the filament's
portion with a static closed coronal structure that led to the
formation of a couple of right-handed and interlocked tori detached
from the Sun and slowly expanding away exactly earthward. The couple
of tori then evolved into a spheromak whose expansion was probably
restrained by an enhanced-density environment due to combined
effects of the neighboring CMEs. The earthward direction of this
ejecta along with its weak expansion and a small mass prevented its
detection with LASCO. Due to the unusually slow expansion, the
disconnected spheromak preserved very strong magnetic field. In
addition, this strong field was pointed almost exactly southward. As
a result, its interaction with the Earth's magnetosphere caused the
surprisingly strong geomagnetic storm on 20 November.

The source region of the compact ICME hitting the Earth obviously
must be close to the solar disk center, which was really the case.
The toroidal magnetic component of the MC was formed from the axial
magnetic field of the eruptive filament. The toroidal magnetic flux
of a perfect spheromak is about 3.5 larger than its poloidal flux.
The poloidal flux of the spheromak was $\approx 5.5 \times
10^{20}$~Mx according to \inlinecite{Moestl2008}, and their estimate
of the complementary magnetic flux of $(11-44) \times 10^{20}$~Mx is
consistent with the expectation for a spheromak, although it was
obtained for a flux rope geometry. The large negative $B_z$ in the
MC was due to its poloidal magnetic field formed from the formerly
static coronal structure, while the original magnetic field of the
filament was mainly responsible for the $B_y$ component, which most
likely was not crucial for the extreme geomagnetic effect (although
$B_y$ is included in Akasofu's $\epsilon$ parameter).

\subsection{Is It Possible to Forecast Such Superstorms?}
The outlined scenario along with complications of the solar event
considered in Papers I\,--\,III leaves a pessimistic impression of
an erratic combination of accidental circumstances that is
impossible to predict. However, there are some promising
circumstances.
\begin{itemize}
 \item
 As the magnetic field extrapolation in Paper~III shows, the
pre-eruption filament was pointed exactly to the coronal null
point. Thus, the topological catastrophe was inevitable and
therefore predictable in principle.
 \item
  An observed manifestation of the topological catastrophe was
the anomalous eruption. Such eruptions are usually visible in the He
{\sc ii} 304~\AA\ line (Paper~I; \opencite{Grechnev2011_AE}) and can
be indicators of potentially dangerous processes on the Sun.
 \item
The compact earthward CME was not detected by LASCO but could be
detected from a vantage point far from the Sun-Earth line. Such a
compact CME/ICME can be probably registered with STEREO imagers.
 \item The 3D reconstructions from SMEI data were very helpful in
recognizing the compact earthward-propagating ICME. The SMEI
observations were terminated in 2011, and the current situation
does not allow such reconstructions to be made in near real-time.
The fruitful employment of the SMEI data confirms that the method
of tomographic reconstructions of heliospheric disturbances from
white-light observations with a wide field of view deserves
further elaboration and implementation in future missions.
\end{itemize}
The major challenge of the 18\,--\,20 November event was its
incomprehensibly large geomagnetic effect. Once a key to the enigma
has been found, investigations into its features could hopefully
make clearer its different manifestations.

\subsection{Concluding Remarks}
\label{S-conclusion}
 Several researchers have contributed to the
study of the 18 November 2003 event and the 20 November magnetic
cloud: \inlinecite{Yermolaev2005}; \inlinecite{Gopal2005c};
\inlinecite{Yurchyshyn2005}; \inlinecite{Ivanov2006};
\inlinecite{Moestl2008}; \inlinecite{Chandra2010};
\inlinecite{Schmieder2011}; \inlinecite{KumarManoUddin2011};
\inlinecite{Marubashi2012}; and others. These efforts certainly have
advanced, step by step, our understanding of solar phenomena and
their space weather outcome.

Various observations have revealed that the magnetic cloud
responsible for the 20 November 2003 superstorm was, most likely, a
compact spheromak disconnected from the Sun. We do not think that
our analysis puts a final point in the study of the 18\,--\,20
November 2003 event. The analysis has revealed several aspects of
the event that were not noticed previously. Our main purpose was to
understand the overall scenario of the event and to answer the major
question why this seemingly ordinary eruptive event gave rise to the
MC with a very large southward magnetic component near Earth, $B_z
\approx -45$ nT, and eventually caused the strongest geomagnetic
storm. In this way our analysis has revealed, for example, an
anomalous eruption with a catastrophe of the filament near a coronal
null point; the transformation of the handedness of a pre-eruption
structure; the necessity to take account of the expansion factor of
an ICME that can be significantly different from its normal
behavior; possible considerable differences of a real magnetic cloud
from an idealized concept, and others. These issues appear to
deserve further investigation, and we hope that our results would
help future studies to address them.

\begin{acks}
We thank V.~Sdobnov, K.~Marubashi, and P.~K.~Manoharan for useful
discussions and the materials, which they kindly made available to
us. We thank the ACE/MAG and SWEPAM teams and the ACE Science Center
for providing the solar wind data. We are grateful to anonymous
reviewers for useful remarks. We thank the Center for Astrophysics
and Space Sciences in USD for heliospheric 3D reconstruction made
from the SMEI mission, which is a joint project of the University of
California at San Diego, Boston College, the University of
Birmingham (UK), and the Air Force Research Laboratory. This study
was supported by the Russian Foundation of Basic Research under
grants 11-02-00757, 11-02-01079, 12-02-00008, 12-02-92692, and
12-02-00037, the Program of basic research of the RAS Presidium
No.~22, and the Russian Ministry of Education and Science under
projects 8407 and 14.518.11.7047.
\end{acks}

\end{article}

\end{document}